\documentclass{article}
\usepackage[T1]{fontenc} % add special characters (e.g., umlaute)
\usepackage[utf8]{inputenc} % set utf-8 as default input encoding
\usepackage{ismir,amsmath,cite,url}
\usepackage{graphicx}
\usepackage{color}
\usepackage{booktabs}
\usepackage{enumitem}
\usepackage{lineno}
\usepackage{caption}
\usepackage{subcaption}
\usepackage{multirow}

\title{YM2413-MDB: A Multi-Instrumental FM Video Game Music Dataset with Emotion Annotations}

\multauthor
{Eunjin Choi$^1$ \hspace{1cm} Yoonjin Chung$^2$ \hspace{1cm} Seolhee Lee$^1$} { \bfseries{JongIk Jeon$^3$ \hspace{1cm} Taegyun Kwon$^1$ \hspace{1cm} Juhan Nam$^1$}\\
 $^1$ Graduate School of Culture Technology, KAIST, South Korea\\
$^2$ Graduate School of AI, KAIST, South Korea\\
$^3$ Department of Industrial Design, KAIST, South Korea\\
{\tt\small \{jech, yoonjin.chung, seolhee\_lee, matji, ilcobo2, juhan.nam\}@kaist.ac.kr}
}

\def\authorname{Eunjin Choi, Yoonjin Chung, Seolhee Lee, JongIk Jeon, Taegyun Kwon, Juhan Nam}

\usepackage[dvipsnames]{xcolor}

\begin{document}

\maketitle
%
% \vspace{-1cm}

\begin{abstract} % 150-200 words
Existing multi-instrumental datasets tend to be biased toward pop and classical music. In addition, they generally lack high-level annotations such as emotion tags.
%In the existing multi-instrumental datasets, genres and instruments are biased toward pop and classical music, and high-level annotations such as emotion tags are rare.
% Existing multi-instrumental datasets lack genre diversity, the number of instruments, and high-level annotations such as emotion. 
In this paper, we propose YM2413-MDB, an 80s FM video game music dataset with multi-label emotion annotations. It includes 669 audio and MIDI files of music from Sega and MSX PC games in the 80s using YM2413, a programmable sound generator based on FM.  %The dataset includes 669 audio and MIDI files of music from Sega and MSX PC games in the 80s using the YM2413 FM sound chip with 19 emotion tags. 
The collected game music is arranged with a subset of 15 monophonic instruments and one drum instrument. They were converted from binary commands of the YM2413 sound chip. Each song was labeled with 19 emotion tags by two annotators and validated by three verifiers to obtain refined tags. We provide the baseline models and results for emotion recognition and emotion-conditioned symbolic music generation using YM2413-MDB.

% Game music takes a large part in shaping the overall atmosphere of a game. However, as the number of game maps and situations increases and characters become more diverse, there is a demand for creating game music more easily and diversely. In addition, game music is suitable for conditional music generation research because the music had clear composition intentions for specific game scenes. However, despite its significance in terms of industry and research, there is a lack of research data to generate game music.
% 
\end{abstract}
\section{Introduction}\label{sec:introduction}
Music generation has been regarded as an attractive research topic for decades. Due to the revival of neural networks, the trend of music generation research has moved from rule-based to data-driven music approaches \cite{lopez2018algoritmic, ji2020comprehensive}. %In data-driven approaches, characteristics of the dataset are becoming important as it often modulates the quality and variance of the generation.
In the data-driven approaches, the dataset determines music styles of the generated output and thus it is desired to have a wide assortment of datasets that cover diverse music genres and styles. However, the majority of music datasets are limited to a single instrument, particularly the piano \cite{pop909-ismir2020, huang2020pop, hung2021emopia, hawthorne2018enabling, kong2020giantmidi}. Multi-instrumental datasets mainly cover popular music \cite{raffel2016learning, sarmento2021dadagp} or classical music \cite{boulanger2012modeling}. In this regard, we introduce a new dataset from 80s FM video game music originally played for YM2413, a programmable sound generator based on FM sound synthesis. We name it the YM2413 Music Database or YM2413-MDB. 

YM2413-MDB provides not only a unique flavor of retro game music but also addresses two important issues in music generation research: the fixed number of instruments and the lack of high-level annotations.  
The most widely used multi-instrumental music dataset is the Lakh MIDI dataset (LMD). While it is the only large-scale MIDI dataset so far, the musical quality of MIDI files is not consistent within the dataset because it is gathered from public sources. Therefore, most studies using LMD generated music with a limited number of instruments for pop music \cite{dong2018musegan, dong2018bmusegan, liang2019midi-sandwich, liang2019midi-sandwich2, hirai2019melody2vec, ren2020popmag, liang2020pirhdy, wu2021musemorphose}. Other datasets, such as JSB Chorale \cite{boulanger2012modeling} and NES-MDB \cite{donahue2018nes}, have only four fixed instruments. Furthermore, the majority of these datasets lack high-level annotations such as genre or mood, which can be used for conditional music generation. Although songs in LMD are matched to entries in the Million Song Dataset (MSD), which has high-level tag annotations, the labels are rather noisy. There exist two symbolic emotion music datasets manually labeled \cite{ferreira2021learning, hung2021emopia}. However, they include only piano music. YM2413-MDB tackles the two issues with the following characteristics.

\begin{itemize}[leftmargin=*]
    \item{\textbf{Multi-instrumental 80s FM Game Music}: The dataset consists of 80's Sega game music played by YM2413, where %YM2413 utilized synthesized sound from Frequency Modulation (FM)
    9 out of 16 instruments can be played at once, which improved the quality of the played game music at that time. 
    From this advancement, the music played by YM2413 contains rich harmonic and rhythmic information. }
    \item{\textbf{Multi-modal}: The dataset contains 669 songs in various formats: binary video game music (VGM) files that contain FM synthesis parameters, converted MIDI files, and rendered audio files which restore the original game music sound. Also, the generated MIDI files can be rendered to audio using an FM-synthesizer plug-in.}
    \item{\textbf{Emotion-annotation}: We annotated the entire dataset with 19 emotion tags in the game context. To make a tag vocabulary with high agreement, each song was annotated by two dedicated annotators and verified by three reviewers. We also confirmed that these tags can be classified using the baseline classification approach.
    }
\end{itemize}

\begin{table*}[]
\centering
% \resizebox{\textwidth{!}{%
\resizebox{0.9\textwidth}{!}{%
\begin{tabular}{l|llcccc}
\toprule
\textbf{Name}              & \textbf{Label Type} & \textbf{Music Style}      & \textbf{\# Inst} & \textbf{\# Avg Tracks} & \textbf{\# Songs} & \textbf{\# Hours}    \\ \midrule
LMD\cite{raffel2016learning}          & None                & General          & 128              & 8.8 / no limit         & 178K              & 9K                   \\
DadaGP\cite{sarmento2021dadagp}                    & None                & Band (Guitar Pro)    & 9                &    3.3 / 9     & 26.2K              & 1.2K                 \\
VGM-RNN\cite{mauthes2018vgm}                    & None                & Game music (NES, crowd source)    & 128                & 6.6 / no limit         & 4.2K              &          101.7*        \\
JSB Chorale\cite{boulanger2012modeling}                    & None                & Classic (Bach Chorale)    & 4                & 4            & 382              & 3.9*                 \\
NES-MDB\cite{donahue2018nes}                    & None                & Game music (early 80s)    & 4                & 2.8 / max 4            & 5.2K              & 46.1                 \\
MOODetector\cite{panda2013multi}                     & 28 adjectives             & General & 128                & 9.2 / no limit                      & 196               & 12.1* \\
VGMIDI\cite{ferreira2021learning}                     & Valence             & Piano-arranged game music & 1                & 1                      & 200               & 6.3* \\
EMOPIA\cite{hung2021emopia}                     & Russell's 4Q        & Pop piano covers          & 1                & 1                      & 1,078             & 11                   \\ \midrule
YM2413-MDB \textbf{(ours)} & 19 adjectives        & Game music (late 80s)     & 16               & 7.2 / max 9            & 669               & 15.4                 \\ \bottomrule

\end{tabular}
}%
% }
\vspace{-0.1cm}
\caption[General Dataset Statistics]{Comparison with existing multi-instrumental or emotion-labeled symbolic music datasets. When the dataset do not limit the instrument, \# Inst is 128. When the dataset has no information of size in hours, it was estimated using the MIDI max tick (denoted as *).}
\label{tab: comparison with other dataset}
 \vspace{-0.2cm}
\end{table*}

% We collected YM2413-MDB manly for the purpose of music generation but it can be also used for various music information retrieval tasks including music emotion recognition. 

We first describe the details of dataset collection and statistical analysis of musical features. Then, we provide baseline results of music emotion recognition in both audio and MIDI domains and emotion-conditioned symbolic music generation using a music generation model. 
%These characteristics of YM2413-MDB can facilitate audio- and symbolic-domain music emotion recognition, emotion-conditioned symbolic music generation, and other emotion-related MIR tasks. 
% which is collected for the purpose of music generation but can be also used for various music information retrieval tasks.
We release YM2413-MDB in the online repository\footnote{https://github.com/jech2/YM2413-MDB}. Examples of generated music are accessible in our demo webpage\footnote{https://jech2.github.io/YM2413-MDB/}.

%%% 아까 지웠던 부분 다시 복구 하였습니다.
% The composition of the entire paper is as follows: We described the details of dataset collection, MIDI conversion, emotion annotation, verification, and dataset representation in Section 3, and the basic midi-level dataset analysis in Section 4. Baseline results of audio- and symbolic-domain music emotion recognition are explained in Section 5, and emotion-conditioned symbolic music generation using GPT2 and Word2Vec embedding of emotion word is presented in Section 6.

\section{Related Work}
\subsection{Characteristics of Game Music}
% 최초의 아케이드 비디오 게임인 <퐁>에서 사용된 단음 효과음을 시작으로, 게임 음악의 발전은 컴퓨터 사운드의 발전과 맥을 같이한다. 기술의 발전에 따라 오늘날에는 다른 미디어 음악과 큰 차이가 없으나, 초창기 비디오 게임들은 몇 가지 특징을 가지고 있다. 
Starting with the single sound effect used in the first video game, \textit{Pong} (1972), the development of game music is in line with the development of computer sound technology \cite{gutierrez2018impact}. Today's game music is not much different from music from other media, such as dramas and movies.
%, similar to the trend of blurring the boundaries between games and other media. 
However, early video game music has a distinct characteristic from other genres of music due to technical constraints at the time. 
%However, early video game music, which had many restrictions, has a distinct characteristic from other genres of music. 
A musicological study on early game music found that some unique composition characteristics are found in the form of arpeggio patterns, compound melodies, and simulated reverb effects using delayed instruments \cite{gutierrez2016programmed}. 
%To analyze early video game music, \cite{gutierrez2016programmed} studied NES game music with a musicological approach. They found some unique composition characteristics for early video game music: arpeggio patterns, 
% which enables perceiving a chord by fast arpeggio playing; 
%compound melodies, 
%which combines the melody and chord in one instrument track;
%and simulated reverb effects using delayed instruments. 
%by delaying the same voice by a dotted eighth note length and playing it with different tracks. 
%These characteristics of retro video game music became lighter as hardware limitations at that time were resolved \cite{gutierrez2018impact}. 
Another characteristic of video game music is that it is composed to be played as a loop \cite{prechtl2016adaptive}. In addition, the transition of playing music occurs when the game context is changed, such as a triggered event: map transition, battle with a boss monster, mission completion, ending, and so on.
%Another characteristic of video game music that is shared with game music used to date will be loop-based \cite{prechtl2016adaptive}. The game music was composed to be played as a loop, and the transition of playing music occurs when the game context is changed, such as a triggered event: map transition, battle with a boss monster, mission complete, ending, etc. To make these transitions more natural, fade-in and fade-out were used. 

\subsection{Game Music Generation}
% 이러한 연구들은 게임음악의 adaptive 함, 혹은 loop적인 성질을 이용하여 게임 음악을 생성하는 시도가 있었으나, 딥러닝의 부상으로 일반적인 음악과 비슷하게 딥러닝을 이용한 연구들로 자연스럽게 shift했다.
%Although music is an important element of gaming that affects mood, 
There are several studies that focus on computer-generated game music. For example, Prechtl used a music generation model based on Markov Chain, which changes the model state in real-time aligned with in-game parameters \cite{prechtl2016adaptive} They tested the model in a live game context.
%Although music is the important element that is in charge of the atmosphere of a game, there are few studies that focus on the game music generation.  To the best of our knowledge, there is one early study focused on adaptive game music. \cite{prechtl2016adaptive} generated game music using classical music and Markov Chain that changed in real-time aligned with the in-game parameters and tested the music in the live game context. 
Mauthes generated video game music using an RNN-based model by collecting video game music crowd-sourced by the game user community \cite{mauthes2018vgm}. Ferreira and Whitehead used piano-arranged video game music to generate music using LSTM and a genetic algorithm with sentiment as a condition  \cite{ferreira2021learning}. 
Donahue et al. generated game music using Transformer-XL\cite{dai2019transformerxl} with the Lakh dataset\cite{raffel2016learning} pre-training and NES-MDB  fine-tuning \cite{donahue2018nes, donahue2019lakhnes}. 

%After the deep learning era, there are a few attempts that utilize neural networks for game music generation. \cite{mauthes2018vgm} generated video game music using an RNN-based approach. They collected video game music that is crowd-sourced by the game user community. \cite{raffel2016learning} used piano-arranged video game music to generate music using LSTM and a genetic algorithm with the sentiment as a condition. 
%Another research that focuses on game music generation is \cite{donahue2019lakhnes}. They generated game music by using Transformer-XL\cite{dai2019transformerxl} with the Lakh dataset\cite{raffel2016learning} pre-training and NES-MDB\cite{donahue2018nes} fine-tuning.

\begin{figure*}[htp!]
 \centerline{
 \includegraphics[width=\textwidth]{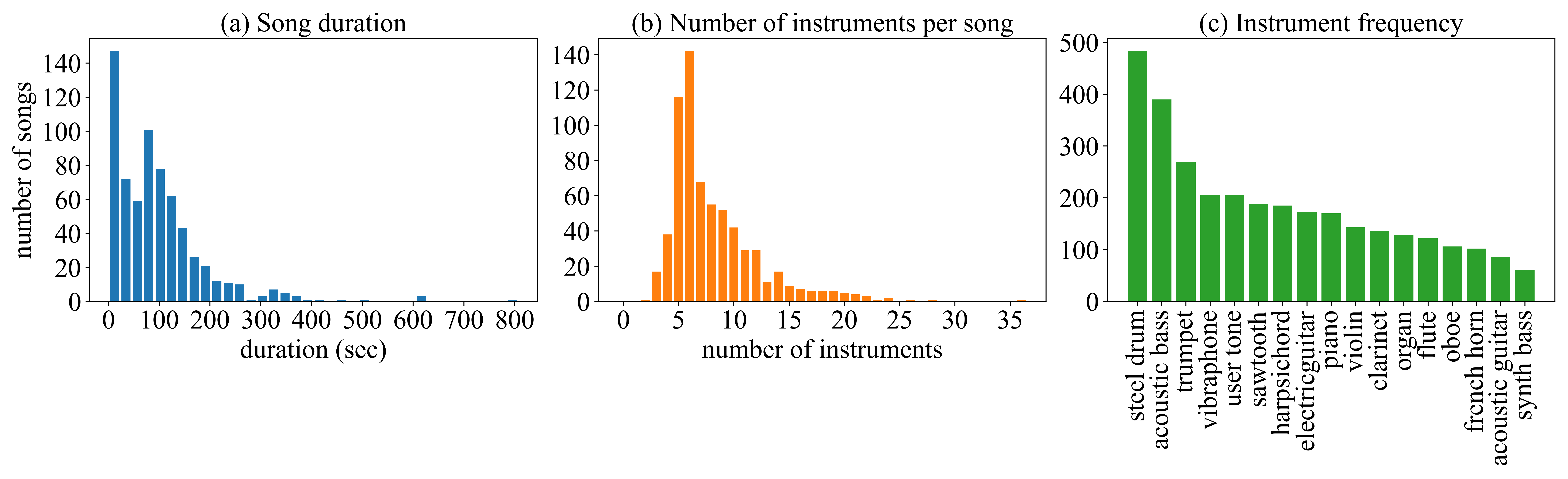}}
 \vspace{-0.5cm}
     \caption{General statistics of YM2413-MDB. (a) Song duration of all songs, (b) Number of instruments per song considering the all program change in single channel, (c) Instrument frequency of whole dataset.}
 \label{fig:general statistics}
 \vspace{-0.2cm}
\end{figure*}
\subsection{Multi-instrumental Symbolic Music Dataset}
% 현재 이용 가능한 멀티 악기 데이터셋으로는 팝, 클래식, 기타, 게임 분야 등이 있다. 가장 크고 잘 쓰였던 게 LMD이다.
% LMD는 워낙에 큰 데이터셋이고, 다양한 악기가 있기 때문에 그래서 최근까지 대부분의 multi-instrumental 연구들이 LMD를 이용하여, 일부 악기만 사용하는 등 변형하여 사용하였다.
% 최근에는 piano쪽에서 giant midi가 나온 것처럼, LMD만큼은 아니지만 DadaGP, Symphony dataset 등 몇 만 개 스케일의 큰 데이터들이 multi-instrument dataset으로 등장하고 있는 추세이다. 
% 멀티 악기 중에서도, 특별히 우리 연구와 가장 비슷한 형태의 데이터로는  NES-MDB가 있다. NES-MDB와의 차이점 설명~
There are a handful of multi-instrumental datasets that cover different music genres, including pop\cite{ panda2013multi, raffel2016learning}, classic\cite{boulanger2012modeling, liu2022symphony}, band\cite{sarmento2021dadagp}, and game\cite{donahue2018nes, mauthes2018vgm} genres. \tabref{tab: comparison with other dataset} compares the datasets to YM2413-MDB. 
Among them, the most widely used one is LMD \cite{raffel2016learning}. Since LMD is a large-scale dataset and has a wide variety of instruments, most multi-instrumental research has used LMD with a set of simplified instruments \cite{dong2018musegan, dong2018bmusegan, liang2019midi-sandwich, liang2019midi-sandwich2, hirai2019melody2vec, ren2020popmag, liang2020pirhdy, wu2021musemorphose} so far. Recently, large-scale datasets with a Band \cite{sarmento2021dadagp} or Orchestra \cite{liu2022symphony} style have been released.
%Recently, similar to GiantMIDI-piano\cite{kong2020giantmidi} from the classical piano music dataset, large-scale data with Band \cite{sarmento2021dadagp} or Orchestra \cite{liu2022symphony} style are released as multi-instrument datasets. 
In the game music genre, NES-MDB, the early 80’s Nintendo Game Music Dataset \cite{donahue2018nes} is most similar to our study. All songs in NES-MDB have four monophonic instruments because of the technical constraints of NES’s Audio Processing Unit. Songs in NES-MDB have a unique style as retro game music and have expressiveness represented with note timings in the precise time units. However, such expressive characteristics require metric analysis such as beat tracking to extract score-level information.  
%Due to such strong constraints, songs in NES-MDB have a unique style that can be captured by deep learning models \cite{donahue2019lakhnes}. Another characteristic of NES-MDB is expressiveness, which represent the timing of notes in precise time units.
%Among the multi-instruments, NES-MDB, the early 80’s Nintendo Game Music Dataset \cite{donahue2018nes} is the most similar type to our study. All music in NES-MDB has four monophonic instruments because of the strong technical constraint of NES’s Audio Processing Unit (APU). Due to such strong constraints, songs in NES-MDB have a unique style that can be captured by deep learning models \cite{donahue2019lakhnes}. Another characteristic of NES-MDB is expressiveness, which represent the timing of notes in precise time units.
%in which the ticks of the MIDI file are set identically with the APU's clock to preserve the original sound. 
% However, the expressive characteristics of NES-MDB makes it difficult to apply further musical analysis or utilize other musical information while generating music. Our YM2413-MDB compensates for the shortcomings of the LMD and NES-MDB; while taking advantage of game music as a dataset, YM2413-MDB has fewer limitations than NES-MDB.

\subsection{Symbolic Music Dataset with Emotion Annotation}
There are a few symbolic music datasets with emotion annotation \cite{ferreira2021learning, hung2021emopia, panda2013multi}. 
%There exist few symbolic music datasets with emotion annotation \cite{ferreira2021learning, hung2021emopia, panda2013multi}. 
MOODetector contains audio, MIDI, and lyrics for 193 music files \cite{panda2013multi}. The music was annotated according to the emotion category used in the MIREX Mood Classification Task \cite{hu2007exploring}. VGMIDI  contains 200 MIDI files of piano-arranged video game music with arousal-valence annotations \cite{ferreira2021learning}. EMOPIA contains 1087 pop piano music with annotations according to Russell’s 4Q \cite{hung2021emopia}. All of these datasets include both MIDI and audio formats.
%Multi-modal MIREX-like emotion dataset version of MOODetector \cite{panda2013multi} contains audio, MIDI, and lyrics of 193 music files. The music was annotated according to the emotion category used in the MIREX Mood Classification Task \cite{hu2007exploring}. VGMIDI \cite{ferreira2021learning} contains 200 MIDI files of piano-arranged video game music with arousal-valence annotations. EMOPIA \cite{hung2021emopia} contains 1087 pop piano music with annotations according to Russell’s 4Q. All of these datasets are multi-modal(MIDI + Audio) datasets.

\section{The YM2413-MDB}
\subsection{Dataset Description}
\label{sec:dataset_description}
The dataset consists of game music played by the YM2413 sound chip. YM2413 is a programmable sound generator based on FM synthesis developed by \textit{Yamaha}. 
%The device is also known as the sound chip of the Japanese version of the \textit{Sega Master System}. 
FM synthesis can generate diverse musical tones with rich harmonics. 
%so that the sound chip can express more diverse instruments than existing sound chips, such as the programmable sound generator (PSG).
YM2413 can play a maximum of 9 instruments simultaneously from 16 instrument presets and one user tone register. The names of the instruments are shown in \figref{fig:general statistics}(c).
%Therefore, YM2413 can play a maximum of 9 instruments simultaneously from 16 instruments preset and one user tone register, which is shown in \figref{fig:general statistics}-(c).
YM2413-MDB contains songs with a short duration for specific game events similar to those in NES-MDB. The statistics are shown in \figref{fig:general statistics}(a). In addition, we observed two unique composition patterns: unison playing and simulated reverb using delay. Unison playing is a pattern of the same melody with single or multiple instruments. This technique renders a rich tone by overlapping the same voices of the FM instrument. The delay pattern is also observed in the NES music to express the reverb by using the delayed same voice  \cite{gutierrez2016programmed}. These patterns show how the composers at that time were concerned about expressing various tones. Therefore, the role and use of each instrument observed in YM2413-MDB seem to be quite different from those in today's composition style.

% This technological advancement increased the general quality of the game music compared to the game music that was played by PSG.: 같은 게임기인데도 japanese version에서 소리 퀄리티가 훨씬 좋음.. 문장 자꾸 동어반복 같아서 뺌

\begin{figure*}[htp!]
 \centerline{
 \includegraphics[width=\textwidth]{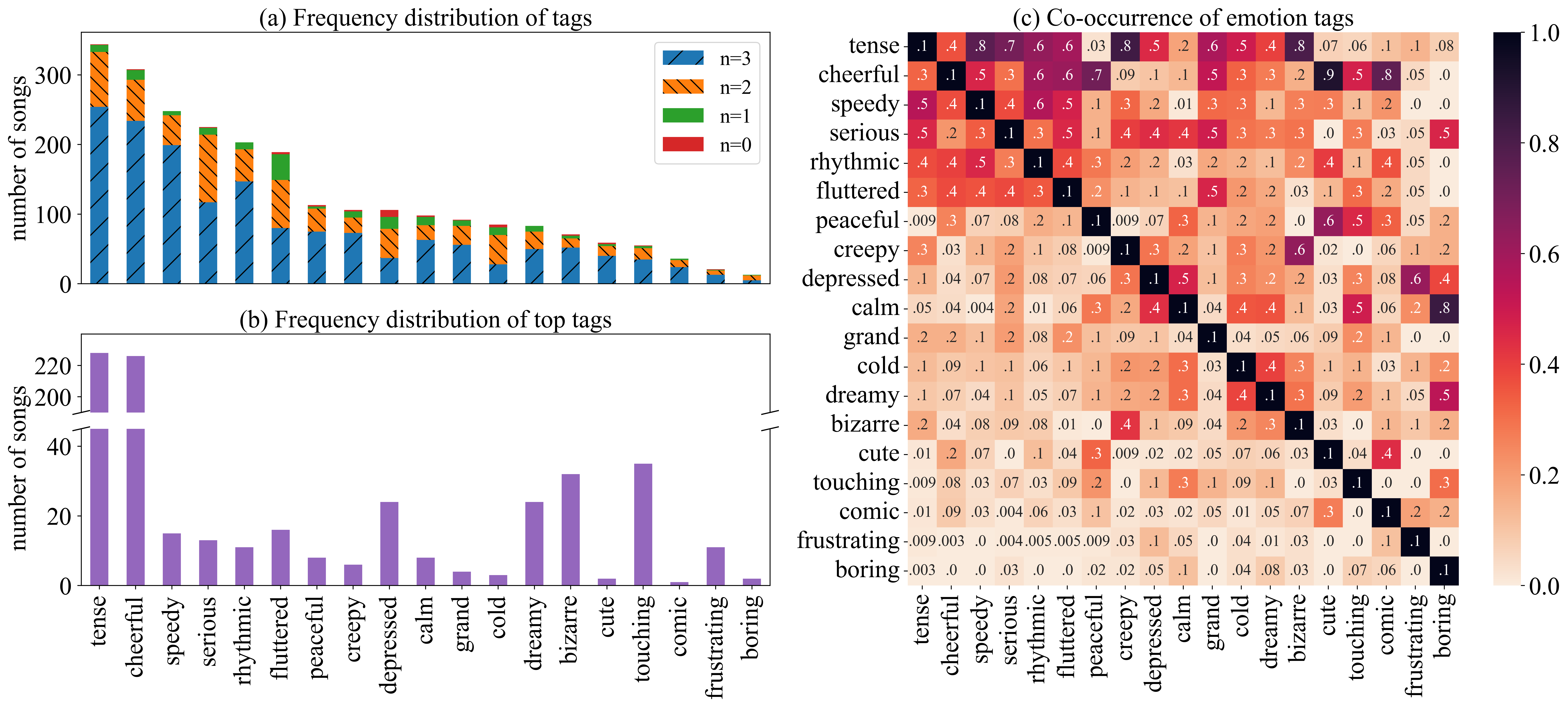}}
 \vspace{-0.2cm}
 \caption{Emotion annotation and verification results. (a) Frequency distribution of emotion tags where n is the number of agreements of verifiers, (b) Frequency distribution of top tags, (c) Co-occurrence of emotion tags. Each column is normalized by the frequency of each tag.}
 \label{fig:emotion tag annotation & verification}
\vspace{-2mm}
\end{figure*}
\subsection{Dataset Collection \& Preprocessing}
We collected Video Game Music (VGM) files from the game music community websites\footnote{https://www.smspower.org/Tags/FM}\footnote{https://vgmrips.net/packs/chip/ym2413}. 
%Video Game Music (VGM) files from ripped game files were collected from the internet sites\footnote{https://www.smspower.org/Tags/FM}\footnote{https://vgmrips.net/packs/chip/ym2413}. 
After data collection, we translated all the game music files to the MIDI format. 
%After data collection, all game music files were converted to MIDI files. 
Since the original VGM files are written as binary command sequences read by the sound chip, the VGM commands are different for all sound chips; for YM2413, there are non-standard commands for time wait, instrument note on/off, program change, volume change, and FM synthesis parameters for one user's custom tone. We conducted the conversion by emulating the YM2413 sound chip following the FM Operator Type-LL (OPLL) Application Manual \cite{YM2413manual}.
%Since the original VGM files consist of binary command sequences that are read by the sound chip register, the VGM commands are different for all sound chips; for YM2413, there are commands of time wait, instrument note on / off, program change, volume change, and FM synthesis parameters for one user's custom tone. MIDI conversion was done by emulating the behavior of the sound chip. 
For the VGM disassembly process and sound chip emulation, we referred to the source code of VGMPlay\footnote{https://github.com/vgmrips/vgmplay}, a software that reads the VGM file and reproduces the sound of hardware sound chips. Also, we exported the VGM files as audio files using this software. 

The original VGM file format expresses all time-related information as ticks with no tempo-related event. In the case of NES-MDB, they converted all MIDI files with 120 beats per minute (BPM) and 22050 ticks per beat to preserve the temporal resolution of 44100Hz without precise tempo calculation. However, the converted MIDI files were not aligned with the metrical structure that standard MIDI files have. This has hindered the use of bar-based music representation, chord recognition, and further music analysis \cite{huang2020pop, hsiao2021compound, ren2020popmag, zeng2021musicbert}.  
%common human-created MIDI files have, which hinders using bar-based music representation such as \cite{huang2020pop, hsiao2021compound, ren2020popmag, zeng2021musicbert}, chord recognition, and further music analysis. 
Therefore, we conducted the metrical alignment using the extracted down-beat information of rendered audio files using Madmom [14]. For each music file, tempo events were added using estimated tempos from all downbeats. Since most of the pieces had a fixed tempo, we used the median value of the beat-wise estimation of tempo. 

In addition, the sound chip synthesizes sounds using frequency values rather than MIDI note numbers; in practice, a frequency is expressed as an F-Number, a discrete value that maps the frequency using 9 bits \cite{YM2413manual}. Therefore, some of the songs in YM2413-MDB were intentionally detuned for the song's ambiance, and instruments had pitch bends and vibratos for more expressiveness. These characteristics make it difficult to map the right MIDI pitch without careful consideration. We treated such cases using the MIDI pitch bend event, such that the integers and decimal points of the calculated MIDI note number are mapped to the MIDI pitch and pitch bend value.

\subsection{Emotion Annotation}
During the annotation and verification process and analysis, we referred to \cite{kim2020semantic} for tag vocabulary refinement, verification procedure, and tag visualization. Two researchers participated in both initial tagging and annotation. Once the dataset was collected, we first listened to the entire game music audio files in the dataset and freely described them with the perceived emotion words. From this process, we found several game-specific emotion words that cannot be directly mapped into both the conventional emotion arousal-valence model and emotion tags used in MIREX music classification\cite{panda2013multi}: fluttered (war-inspiring), grand, bizarre, frustrating, comic, faint, and touching. As discussed in \cite{gomez2021music}, the categorical approach is suitable for these complex emotions rather than the dimensional approach. Therefore, we decided to annotate the emotions as word tags. 

From the emotion descriptions of all songs in the dataset, we obtained 35 emotion-related tags. We grouped similar emotion words according to a Korean crowdsource BGM repository \footnote{https://bgmstore.net/} 
and finally obtained 19 emotion tags. After the tag vocabulary was refined, we conducted the labeling. While annotating the dataset, each of the annotators selected one dominant emotion tag for each music which we call ``top tag'' in this paper. Among the top tag annotated by the two annotators, the final top tag was selected as having a higher agreement after verification.

\subsection{Emotion Verification}
After the annotation step, each music file was validated by three verifiers; in total, ten people participated as verifier. During verification, verifiers were asked to mark the emotion tag if they disagree with the annotation. We provided verifiers with the annotator's description of each emotion tag to reduce the ambiguity of each emotion tag. After verification, tags with low agreement (lower than two) were excluded from the annotation of each music. When the top tag was excluded after verification (26 songs), the tag with the highest agreement was chosen as the new top tag.

\figref{fig:emotion tag annotation & verification}(a) shows the overall emotion tag distribution of the YM2413-MDB after verification. Also, the agreement portion of verification explains the subjectivity of each tag. In particular, `serious', `fluttered', `depressed', and `cold' tags have less agreement than other tags. The frequency distribution of top tags is shown in \figref{fig:emotion tag annotation & verification}(b), in which the imbalance is more noticeable than the frequency distribution of each tag. In \figref{fig:emotion tag annotation & verification}(c), we investigated the independence of each tag by calculating the co-occurrence of emotion tags. The words that often appear together(>0.8) are as follows: `speedy'-`tense', `creepy'-`tense', `bizarre'-`tense', `cute'-`cheerful', `comic'-`cheerful', and `boring'-`calm'. 

% All participants were native Korean aged from twenty-four to thirty-one, with at least one retro video game-playing experience.

Annotation and verification for the Korean tags were operated since all participants were native Korean. They are aged from twenty-four to thirty-one, with at least one retro video game-playing experience.

\begin{figure*}[!htp]
 \centerline{
 \includegraphics[width=\textwidth]{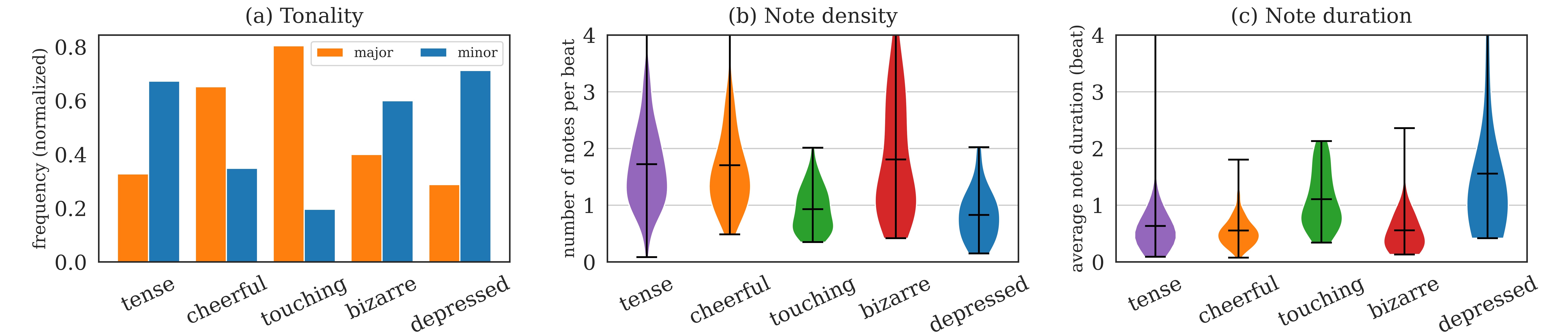}}
 \vspace{-0.2cm}
 \caption{Data distribution for different top tag classes. (a) Normalized frequency of tonality histogram, (b) Note density, (c) Note duration. The outlier sample points were omitted for clear visualization.}
 \label{fig:data analysis plot}
  \vspace{-0.2cm}

\end{figure*}   

\subsection{Data Representation}
% 데이터셋의 특성 상 한 번에 여러 악기를 표현 가능한 것이 필요하였고 임의의 악기 조합이 표현 가능한 멀티 악기 표현 방식 중에서 MMM을 선택했다.
For the baseline symbolic-domain classification and generation, we used the event-based representation called Multi-Track Music Machine (MMM) \cite{ens2020mmm}. The MMM representation can handle arbitrary instrument combinations. As discussed in Section \ref{sec:dataset_description}, our dataset contains unison patterns; the same instrument appears in multiple tracks. To apply representations suggested for multi-instrumental music that assume each instrument appears at most once \cite{donahue2019lakhnes, ren2020popmag, zeng2021musicbert}, we should discard instrument tracks that appear multiple times in a song. However, we did not apply the strategy due to the small dataset size. In our preliminary study, experiments with multi-instrumental representation suggested by \cite{donahue2019lakhnes} only generate a drum with a large number of notes. We speculate that it was due to the frequency imbalance of tokens; since the representation suggested by \cite{donahue2019lakhnes} combines musical instruments and pitch tokens, the number of tokens increases in proportion to the number of instruments, making the model difficult to learn pitch information.
Therefore, we did not use those non-MMM style representations.

To re-implement MMM, we calculated the note density for each instrument by dividing the total number of note counts by the active bar number, which is the number of bars in which at least one note is played. Also, we quantized the music samples with 48 grids per bar. 
After removing the bar-fill-related vocabulary of MMM which is not necessary for our tasks, 
%Since our paper does not cover inpainting tasks, we did not use the Bar-fill related vocabulary of MMM. 
%Therefore, 
we used a total of 447 tokens: five tokens for the piece, track, and bar; 48 tokens for time delta tokens; 128 tokens for each instrument, note on, and note off. Since the MMM representation can express a fixed number of bars, we used four bars with a maximum of six instruments, which the transformer attention window size of 1024 can handle. Fixing the length of generated music is suitable for game music because the game music is loop-based, as discussed in Section 2. We sampled the instrument when the number of instruments was greater than six. Due to the sparsity issue, we adapted the dataset for training. The instruments were mapped into six categories during training: piano, string, woodwind, brass, guitar, bass, and drum. Tracks that change within the same channel are all adapted to be treated as one instrument.

\section{Dataset Analysis}
Emotion is associated with various musical features such as harmony, rhythm, and loudness \cite{9229494}. 
%It is known that there is a correlation between emotion and various musical features such as harmony, rhythm, and loudness \cite{9229494}. 
We investigated how the musical elements extracted from YM2413-MDB are correlated with the collected emotion tags. We used note density, note duration, and tonality following the previous work \cite{hung2021emopia}. Due to the imbalanced distribution of top tags, we focused on songs with five top tags: `tense', `cheerful', `touching', `bizarre', and `depressed'. When extracting the features, we excluded drum tracks to consider the pitch information only when estimating tonality and fairly compared songs with and without drums. 

\subsection{Tonality}
The major/minor tonality is generally connected to positive/negative emotional states \cite{10.2307/40285496}.  
%Musical psychology studies showed that the major/minor tonalities are associated with positive/negative emotional states. 
We measured tonality using the Krumhansl-Schmuckler key-finding algorithm \cite{10.2307/40285812} implemented in music21 \cite{cuthbert2010music21}. 
%The tonalities were measured from the musical keys using the Krumhansl-Schmuckler Key-Finding Algorithm \cite{10.2307/40285812} implemented in music21 \cite{cuthbert2010music21}. 
\figref{fig:data analysis plot}(a) shows a clear trend that the major tonality is frequently used for positive emotions (e.g. `cheerful', `touching'), and the minor tonality is frequently used for negative emotions (e.g. `tense', `bizarre', `depressed').

\subsection{Note Density and Duration}
Another musical feature related to emotion is rhythm \cite{schellenberg2000perceiving}. We use note density and note duration as indirect cues to extract the rhythmic characteristics. The note density refers to the number of notes per beat, which was obtained by dividing the total number of notes by the total number of beats. It was normalized by the number of instruments per song. The note duration was calculated by averaging the note duration value in beat units. As shown in \figref{fig:data analysis plot}(b-c), songs with `tense', `cheerful', and `bizarre' tags showed higher note density than the `touching' and `depressed' tags. Most songs with `tense', `cheerful', and `bizarre' tags consisted of notes shorter than one beat, whereas songs with `touching' and `depressed' tags were widely distributed with longer notes. We also measured the note velocity, but all tag groups maintained a similar distribution. It is because songs in YM2413-MDB do not utilize the velocity features well; in most cases, the velocity of notes is changed together only when program change events occur.

\begin{table}[tp]
 \begin{center}
\resizebox{\columnwidth}{!}{%
    \begin{tabular}{l|ccc|cc}
    \toprule
    \multirow{2}{*}{\textbf{Model}} & \multicolumn{3}{c|}{\textbf{Emotion}} & \multicolumn{2}{c}{\textbf{Top Tag}} \\ \cline{2-6} & \textbf{4Q}  
    & \textbf{Arousal}    & \textbf{Valence}& \textbf{Tense}    & \textbf{Cheerful}    \\ \midrule
    Logistic regression    & 0.48 & 0.76 & 0.56 & 0.66 & 0.66 \\ \midrule
   % \begin{tabular}[c]{@{}l@{}}Logistic \\ regression\end{tabular}    & 0.48 & 0.76 & 0.56 & 0.66 & 0.66 \\ \midrule
   LSTM-Attn\cite{lin2017structured} + MMM\cite{ens2020mmm}& 0.44 & 0.69 & 0.59 & 0.68 & 0.70  \\ \bottomrule
   % \begin{tabular}[c]{@{}l@{}} LSTM-Attn\cite{lin2017structured}\\+ MMM\cite{ens2020mmm}\end{tabular}       & 0.44 & 0.69 & 0.59 & 0.68 & 0.70  \\ \bottomrule
    \end{tabular}
    }%{'AV': 0.48, 'A': 0.76, 'V': 0.56}
\end{center}
\vspace{-0.5cm}
 \caption{Symbolic-domain classification accuracy. 4Q indicates Russell's four quadrants.}
 \label{tab:symbolic domain classification}
\end{table}

\begin{table}[tp]
\vspace{-0.2cm}
 \begin{center}
\resizebox{\columnwidth}{!}{%
    \begin{tabular}{l|ccc|cc}
    \toprule
    \multirow{2}{*}{\textbf{Model}} & \multicolumn{3}{c|}{\textbf{Emotion}} & \multicolumn{2}{c}{\textbf{Top Tag}} \\ \cline{2-6} & \textbf{4Q}  
    & \textbf{Arousal}    & \textbf{Valence}& \textbf{Tense}    & \textbf{Cheerful}    \\ \midrule
    Logistic regression & 0.38 & 0.72 & 0.54 & 0.60 & 0.66 \\ \midrule
    Short-chunk ResNet\cite{won2020evaluation}       & 0.65 & 0.78 & 0.60 & 0.69 & 0.65  \\ \bottomrule
    \end{tabular}
    } %{'AV': 0.38, 'A': 0.72, 'V': 0.54}
\end{center}
\vspace{-0.5cm}
 \caption{Audio-domain classification accuracy. 4Q indicates Russell's four quadrants.}
 \label{tab:audio-domain classification}
 \vspace{-0.5cm}
\end{table}

\section{Music Emotion Recognition}
We provide our baseline symbolic and audio-domain emotion recognition results using YM2413-MDB. For both symbolic- and audio-domain classification, we used the source code and experimental setting (4Q, Arousal, Valence) of \cite{hung2021emopia} and adapted it to our dataset configuration. We used the same model for classification (logistic regression, LSTM-Attn \cite{lin2017structured}, and short-chunk ResNet \cite{won2020evaluation}). For symbolic domain classification, we used a re-implemented MMM representation with 8 bars and 6 instrument chunks. Due to the small volume with imbalanced emotion tags, it was difficult to use the original emotion tags for emotion recognition directly. Instead, we mapped the top tags into Russell's four quadrants and classified a song into one of the four categories; for tags that do not appear in Russell's circumplex model, we manually mapped to one quadrant of the most similar emotion according to the subjectivity of the annotator\footnote{https://github.com/jech2/YM2413-MDB/blob/main/emo4Qmap.png}. We also provide the baseline binary classification results using `tense' and `cheerful' top tags. Other tags showed low classification results due to severe imbalance. We split the dataset with an 8:1:1 ratio for train, validation, and test via stratified sampling. 

The emotion classification results of symbolic- and audio-domain are shown in \tabref{tab:symbolic domain classification} and \tabref{tab:audio-domain classification}. Both symbolic- and audio-domain classification results of top tags show that these emotions can be recognized by the baseline models. Since the dataset is multi-instrumental and the class imbalance is more severe than EMOPIA, the baseline classification was lower than those reported in \cite{hung2021emopia}. Also, we note that the symbolic-domain classification performance of the LSTM-Attention model with MMM was poorer than that of the logistic regression. We conjectured that it was difficult to learn the temporal relations between instruments when using MMM representation.

\section{Emotion-Conditioned Symbolic Music Generation}
\subsection{Data Preparation}
We provide the baseline emotion-conditioned symbolic music generation results using YM2413-MDB. Due to the emotion tag class imbalance, emotion conditioning was ineffective when using the entire set of emotion top tags. Instead, we selected `cheerful' and `depressed' as representative emotion conditions that showed contrasting features in the dataset analysis (Fig ~\ref{fig:data analysis plot}) and conducted conditional generation using the data with the two tags. Similar to \cite{donahue2019lakhnes}, as our dataset size was small to train the deep learning model from scratch, we used LMD for pre-training the music generation model and fine-tuned it using YM2413-MDB. We filtered too short MIDI files which are less than 3 secs. We also filtered short samples annotated as sound effects. After filtering, we used 10K samples for validation and test sets when we trained the LMD. A total of 286 songs were used for fine-tuning, and four songs containing both `cheerful' and `depressed' tags were excluded. 

\subsection{Model \& Train Setting}
% Language modeling으로 music generation 하는 많은 연구들에서 Transformer-XL을 사용하고 있다. 하지만, impirically, Transformer-XL 보다 MMM에서 사용하였던 GPT2가 생성 샘플의 퀄리티가 더 좋아서 GPT2를 사용하게 되었다. 또한, MMM을 reproduce할 때 layer 수를 늘려주니 lakh pre-train 생성결과가 더 좋았다. 그래서 우리는 학습에 해당 모델을 사용하게 되었다.
Transformer-XL\cite{dai2019transformerxl} is widely used in music generation studies with language modeling approach \cite{donahue2019lakhnes, huang2020pop, wu2020jazz, sarmento2021dadagp}. GPT2\cite{radford2019language} is a language model that showed high fine-tuning performance in NLP without changing the model structure and was also used in several music generation studies \cite{ens2020mmm, musenet-2019}. In our preliminary study, we observed that GPT2 generated better quality samples than Transformer-XL. In addition, the generation samples from lakh pre-train were better when adding more layers than \cite{ens2020mmm} used (six layers). 

Therefore, we used the 12-layer GPT2\cite{radford2019language} with an attention window size of 1024 for our baseline generation model. Before passing the main architecture, each token of event sequence input was embedded as 512 dimensions. To maintain the dimension while calculating attention, the number of attention heads and attention head dimensions were set to 8 and 64, respectively. For all layers, the dimension of the feed-forward layer after the attention module was 1024. The initial learning rate was set to 5e-5, and we used the AdamW optimizer with a linear scheduler. To generate music with emotion condition, we fine-tuned the pre-trained GPT2 model using YM2413-MDB.
%For fine-tuning, we freeze the first 8 encoder layers and fine-tune the rest.

Also, by using the in-attention mechanism proposed in \cite{wu2021musemorphose}, we projected the emotion tag embedding to the same space as the hidden states, then summed the projected condition with the hidden states at each self-attention layer. For emotion tag embedding, we used the pre-trained Word2Vec embedding \cite{mikolov2013w2v} of `cheerful' and `depressed' tags, which were also updated during fine-tuning.

\subsection{Generated Results}
\begin{figure}[]
 \centerline{
 \includegraphics[width=\columnwidth]{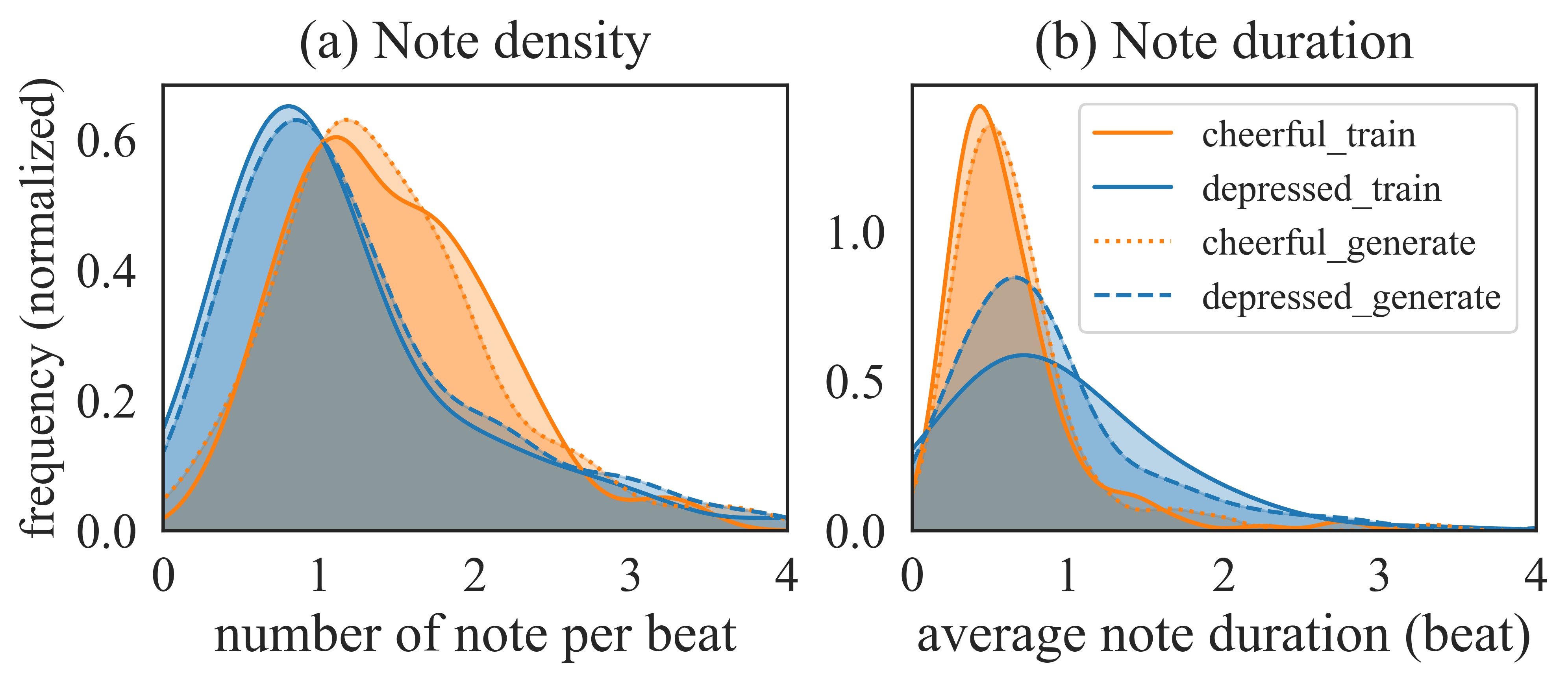}}
 \caption{The note density and note duration distribution of the train set and generated results when emotion condition contains `cheerful' and `depressed'. Each histogram is smoothed using kernel density estimation.}
 \vspace{-0.5cm}
 \label{fig: cheerful & depressed objective histogram}
\end{figure}

Although we provided the baseline classification results, the performance was not sufficient to evaluate conditional music generation. Therefore, in this section, we show the overall tendency of the generation model through the objective measure. To see the validity of emotion conditioning in music generation, we compared 100 model-generated songs conditioned on `cheerful' and `depressed' emotions. We compared the histogram of note density and note duration of the generated results with the model inputs. For model inputs, the first 4 bars of each of the 284 songs were analyzed. As shown in \figref{fig: cheerful & depressed objective histogram}, given `depressed' conditions, the model generates samples with a longer note duration and lower note density than when using `cheerful' condition. However, when we listened to the generated samples, we found some of the generated samples struggled with temporal information, such as note onset time. It seems delayed notes in the dataset hindered learning the overall beat structure of the music. Although the model inputs had distinct distributions for major/minor tonalities, the generated results did not reflect these features. We suppose that it is due to the characteristics of game music that we observed in YM2413-MDB; the appearance of whole-tone or chromatic patterns makes it difficult for the model to learn tonality.

% When we investigated the music features of the YM2413-MDB, the pitch histogram showed no significant difference according to emotion tags, unlike major/minor keys. The musical key is a high-level feature that varies based on how the notes are arranged, making it challenging for the model to understand.

\section{Conclusion}
% one to two paragraphs 
In this paper, we introduced YM2413-MDB, a multi-instrumental FM video game music dataset annotated with emotion tags. Using this dataset, we conducted a basic statistical analysis of musical features and provided the baseline results for emotion recognition and conditional music generation. 
%However, it seems to be one of the big limitations of this study is that emotion annotation and verification were dealt with only Korean subjects. In addition, our paper brings up the question of how to aggregate multi-labeled motion, which needs further study.
%Furthermore, 
We plan to study retro game music compositional styles further and improve music emotion classification and emotion-conditioned music generation. Also, we will investigate other uses of YM2413-MDB such as multi-instrumental music transcription, source separation, and structure analysis.
%do a musicological analysis of emotion and game music composition, improve music emotion classification, and emotion-conditioned music generation, and investigate of the potential of YM2413-MDB to other MIR tasks such as multi-instrumental music transcription, source separation, and structure analysis.

% 혹시 지금의 청취평가의 한계를 이야기하며 추후에 인게임에서 테스트 등도 필요할 것 같다 이런 문장도 필요할지요..!

% For bibtex users:
\section{Acknowledgment}
This work was supported by Year 2022 Culture Technology R\&D Program by the Ministry of Culture, Sports and Tourism and Korea Creative Content Agency (Project Name: Research Talent Training Program for Emerging Technologies in Games, Project Number: R2020040211) and Institute of Information \& communications Technology Planning \& Evaluation (IITP) grant funded by the Korea government (MSIT) (No.2019-0-00075, Artificial Intelligence Graduate School Program (KAIST)).

\bibliography{ISMIRtemplate}

\end{document}